\documentclass[letterpaper]{article}

\usepackage{natbib,alifeconf}  %% The order is important
\usepackage{csquotes}
\usepackage{amsmath}
\usepackage{subcaption}
\usepackage{booktabs}
\usepackage{xcolor}
\usepackage{mathtools}

\newcommand{\eqnref}[1]  {equation~(\ref{#1})}
\newcommand{\figref}[1]  {Fig.~\ref{#1}}
\newcommand{\tabref}[1]  {table~\ref{#1}}

\title{Predictions in the eye of the beholder: an active inference account of \\Watt governors} %(or ``Do Watt-governors dream of steam engines?'', ``Do Watt-governors believe in their steam engines?'', ``Predictions in the eye of the beholder: an active inference account of Watt governors'', or ````Evidence'' for predictive coding in steam engines regulators'', or ``Predictive coding, here, there, and everywhere'')}
\author{Manuel Baltieri$^{1,2}$, Christopher L. Buckley$^{2}$, Jelle Bruineberg$^{3}$ \\
\mbox{}\\
$^1$ Laboratory for Neural Computation and Adaptation, RIKEN Centre for Brain Science, Wako City, Japan \\
$^2$ Evolutionary and Adaptive Systems Research Group, Department of Informatics, University of Sussex, Brighton, UK \\\
$^3$ Department of Philosophy, Macquarie University, Sydney, Australia \\
manuel.baltieri@riken.jp} % email of corresponding author

\begin{document}
\maketitle

\begin{abstract}
% Abstract length should not exceed 250 words
% \textbf{A generative model describing an imaginary engine from the perspective of the governor, except this is the generative model from the perspective of the observer ;). The imaginary engine observes the angle of the arms of the governor and regulates its valve based on such observations. Who's the observer? Is the governor observing the engine? The other way around? Is the experimenter observing the coupled system?}

Active inference introduces a theory describing action-perception loops via the minimisation of variational (and expected) free energy or, under simplifying assumptions, (weighted) prediction error. Recently, active inference has been proposed as part of a new and unifying framework in the cognitive sciences: predictive processing. Predictive processing is often associated with traditional computational theories of the mind, strongly relying on internal representations presented in the form of generative models thought to explain different functions of living and cognitive systems. In this work, we introduce an active inference formulation of the Watt centrifugal governor, a system often portrayed as the canonical ``anti-representational'' metaphor for cognition. We identify a generative model of a steam engine for the governor, and derive a set of equations describing ``perception'' and ``action'' processes as a form of prediction error minimisation. In doing so, we firstly challenge the idea of generative models as \emph{explicit} internal representations for cognitive systems, suggesting that such models serve only as \emph{implicit} descriptions for an observer. Secondly, we consider current proposals of predictive processing as a theory of cognition, focusing on some of its potential shortcomings and in particular on the idea that virtually any system admits a description in terms of prediction error minimisation, suggesting that this theory may offer limited explanatory power for cognitive systems. Finally, as a silver lining we emphasise the instrumental role this framework can nonetheless play as a mathematical tool for modelling cognitive architectures interpreted in terms of Bayesian (active) inference.
\end{abstract}

\section{Introduction}
The free energy principle (FEP) has been proposed as a framework to study perception, action and higher order cognitive functions using probabilistic generative models \citep{Friston2010biocyb, hohwy2013predictive, clark2015surfing, buckley2017free}. Under the FEP and related approaches, including the Bayesian brain hypothesis, perception is usually characterised as a process of (approximate Bayesian) inference on the hidden states and causes that generate sensory input. Predictive coding models describe how this process may be implemented in a biologically plausible fashion by minimising a mismatch error between incoming sensations and predictions, or rather estimates, of these sensations produced by a probabilistic generative model \citep{rao1999predictive, spratling2016predictive}. Active inference extends this account of perceptual processes by 1) noting that they can be treated as a special case of a more general framework based on the minimisation of variational free energy under Gaussian assumptions and by 2) proposing a description of action and behaviour consistent with the minimisation of prediction error and variational (and expected) free energy \citep{Friston2010biocyb, friston2017active}. Active inference thus proposes that agents minimise prediction errors by both generating better estimates of current and future sensory input and, at the same time, acting in the environment to directly update this input to better fit current predictions. These actions are biased towards normative constraints (in the form of prior Bayesian beliefs) that ensure their very existence, closing the sensorimotor loop and sidestepping ``dark room'' paradoxes \citep{Friston2010biocyb, friston2012dark, buckley2017free, friston2017active, baltieri2019dark}.

In (philosophy of) cognitive science, active inference is usually identified within the framework of \emph{predictive processing} \citep{clark2013whatever}. Predictive processing and active inference are often thought to align with more representationalist views of cognition \citep{froese2013brain, gladziejewski2018predictive}. One part of the cognitive science community sees this as a possible advantage \citep{hohwy2013predictive, wiese2017vanilla, gladziejewski2018predictive} while others see it as one of its main drawbacks \citep{froese2013brain, anderson2017bayes, zahavi2017brain}. Others have argued it may be consistent with embodied and enactive perspectives of cognition, claiming that the strengths of the FEP reside in generative models with no explicit representational role \citep{bruineberg2018anticipating, kirchhoff2017there}. A different perspective highlights the potential of the FEP for the formalisation of ``action-oriented'' views of cognition \citep{engel2016pragmatic, clark2015surfing}, attempting to reconcile computational views and embodied/enactive positions.

As a thinking tool on the role of predictive processing as a theory of cognition, we introduce an active inference re-interpretation of a now classical example in the literature of ``anti-representational'' accounts of cognitive systems, the Watt governor. The Watt (flyball or centrifugal) steam governor was used by \cite{van1995might, van1998dynamical} as a paradigmatic system to challenge the dominant cognitivist understanding of cognition. \cite{van1995might, van1998dynamical} claimed that dynamical systems theory provided a better language to explain the inner workings of cognitive agents, similarly to what engineers have done with the Watt governor in terms of attractors, stability analysis, etc.. At the same time, he then questioned whether computational descriptions of the governor offered any explanatory power -- over -- the use of dynamical systems theory. While many proponents of the dynamicist view see the computational metaphor as superfluous in many cases \citep{chemero2009radical}, or intimately tied to a deeply flawed philosophy of mind \citep{dreyfus1972computers}, others see computational and information-theoretic descriptions of a system as useful epistemic tools offering interpretations that are complementary to a dynamical systems analysis \citep{bechtel1998representations, beer2015information}. In both cases, the nature of the tight coupling between a flyball governor and its steam engine is accepted, and the resulting mode of cognition departs from the cognitivist one,  i.e., a governor does not ``read'' the speed of the engine to ``compute'', offline, the next best action. In this light, a more appropriate explanation of such coupled systems mandates circular, rather than linear causality, in line with embodied/enactive approaches to cognitive science highlighting the importance of studying the dynamical interaction of an agent and its environment.

\section{The Watt governor}
\begin{figure}[ht!]
  \centering
  \includegraphics[width=.8\linewidth]{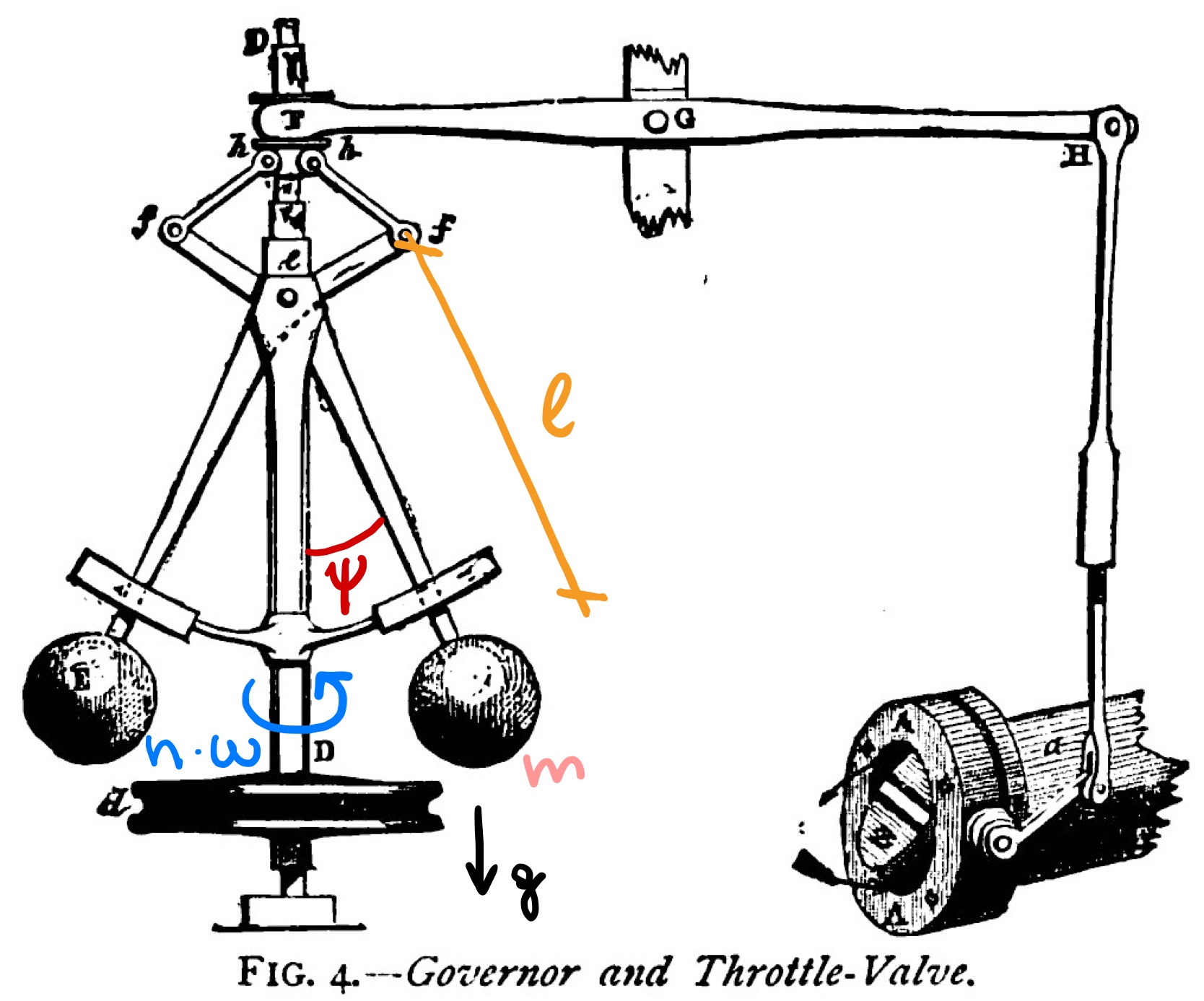}
  \caption{\textbf{The Watt Governor.} The Watt, or centrifugal, governor connected to a throttle valve regulating the flow of steam allowed into the cylinders of a steam engine. (Original image courtesy of Wikimedia Commons.)}
  \label{fig:CentrifugalGovernor}
\end{figure}

The centrifugal (or flyball) governor was introduced as a mechanism to harvest steam power for industrial applications, controlling the speed of steam engines using properties of negative feedback loops \citep{aastrom2010feedback}. The centrifugal governor regulates the amount of steam admitted into a cylinder via a mechanism that opens and closes a valve controlling the amount of steam released by an engine. This regulation requires balancing the forces applied to a pair of flyballs secured via two arms to a rotating spindle, geared to a flywheel driven by a steam engine (\figref{fig:CentrifugalGovernor}). 

At rest, the flyballs are subject to gravitational force and the two arms are in a vertical position while the engine's valve is fully open. As the engine is powered and steam flows into the cylinders via the fully open valve, the engine's flywheel velocity is increased, alongside the vertical spindle's angular velocity. The attached flyballs' kinetic energy thus also increases, counteracting the effects of the gravitational force, slowly bringing the arms away from a vertical position. Once the spindle's velocity reaches a certain bound set by the physical properties of the system (and thus indirectly by the engineer who built it), the steam valve of the engine is slowly closed via a beam linkage connected to a thrust bearing attached to the flyballs' arms. When the steam flow decreases, the vertical spindle slows down, reopening the valve that will move the flyballs to increase once more the speed of the shaft, closing the valve, etc. until a stable equilibrium is reached for a certain steam flow associated to a specific angle between the flyballs' arms and the vertical shaft.

Using a standard formulation by \cite{pontryagin1962ordinary}, based on previous work by \cite{vyshnegradsky1877direct} and \cite{maxwell1868governors}, we define a Watt governor as a conical pendulum with two flyballs (the ``bobs'') travelling at constant angular speed $\phi$
\begin{align}
    \ddot{\psi} & = (\phi)^2 \sin (\psi) \cos (\psi) - \frac{g}{l} \sin (\psi) - \frac{b}{m} \dot{\psi}
    \label{eq:Governor}
\end{align}
based on a simple derivation of Newtonian's laws from \figref{fig:CentrifugalGovernor}, and with all variables explained in more detail in \tabref{tab:variablesWattGovernor}.
\begin{table}[h]
\centering
\caption[Watt governor, variables and constants.]{
{Watt governor, variables and constants.}}
    \begin{tabular}{cl}
        \toprule
        \bf  & \bf Description \\ \midrule
        \( \psi, \phi \) & Flyball arm angle and velocity ($\dot{\psi} = \phi$)\\ 
        % \( \theta \) & shaft angular displacement \\
        % \( \phi \) & Governor flywheel angular speed ($\dot{\psi} = \phi$) \\
        \( \omega \) & Steam engine flywheel angular speed \\% ($\dot{\theta} = \omega$) \\
        \( I \) & Steam engine flywheel moment of inertia \\
        \( G \) & Torque induced by engine load \\
        \( n \) & Gear or transmission ratio \\
        \( g \) & Gravitational acceleration \\
        \( l \) & Length of flyball arms \\
        \( b \) & Friction constant \\
        \( m \) & Flyball mass \\
        \( k \) & Constant relating flyball height and engine torque \\ \bottomrule
    \end{tabular}
\label{tab:variablesWattGovernor}
\end{table}
%
% \begin{table}[h]
% \centering
% \caption[Watt governor, variables and constants.]{
% {Watt governor, variables and constants.}}
%     \begin{tabular}{ccc}
%         \toprule
%         \bf  & \bf Description & \bf Value \\ \midrule
%         \( \psi \) & flyball arm angle & -\\ 
%         \( \theta \) & shaft angular displacement & - \\
%         \( \phi \) & governor flywheel angular speed ($\dot{\psi} = \phi$) & - \\
%         \( \omega \) & steam engine shaft angular speed ($\dot{\theta} = \omega$) & - \\
%         \( I \) & steam engine flywheel moment of inertia & - \\
%         \( G \) & torque by engine load & - \\
%         \( n \) & gear or transmission ration & - \\ 
%         \( g \) & gravitational acceleration & \(9.81\)~m/s \(^2\) \\ 
%         \( l \) & length of flyball arm & \(1\) m \\ 
%         \( b \) & friction constant & \(2\) m/s \\
%         \( m \) & flyball mass & \(1\) kg \\
%         \( k \) & flyball height to engine torque constant & \(1\) m/s \\ \bottomrule
%     \end{tabular}
% \label{tab:variablesWattGovernor}
% \end{table}
%
A steam engine is then attached via a flywheel with angular speed $\omega$, moment of inertia $I$ and a load torque $G$
\begin{align}
    I \dot{\omega} & = k \cos (\psi) - G
    \label{eq:Engine}
\end{align}
The flywheel is geared to the vertical spindle so that the angular velocities of the engine's flywheel and the vertical spindle are proportional, $\phi = n \omega$, up to a constant $n$, the gear ratio. With these assumptions, we can reduce the system to a set of first order coupled differential equations,
\begin{align}
    \dot{\psi} & = \phi \nonumber \\
    \dot{\phi} & = (n \omega)^2 \sin (\psi) \cos (\psi) - \frac{g}{l} \sin (\psi) - \frac{b}{m} \dot{\psi} \nonumber \\
    \dot{\omega} & = \frac{k}{I} \cos (\psi) - \frac{G}{I}
    \label{eq:WattGovernorFirstOrder}
\end{align}
We then find the equilibrium of this system by equating to zero the left-hand side of \eqnref{eq:WattGovernorFirstOrder} and by defining a constant shaft velocity $\omega_0$ and fixed arm angle $\psi_0$ where the arm angular velocity is zero, $\phi = \phi_0 = 0$,
\begin{align}
    \phi_0 & = 0 \nonumber \\
    \cos(\psi_0) & = \frac{G}{k} \nonumber \\
    n^2 \omega_0^2 & = \frac{g}{l \cdot \cos(\psi_0)}
    \label{eq:WattGovernorFirstOrderEquilibrium}
\end{align}
The system is henceforth linearised near its equilibrium to simplify the analysis, see \cite{maxwell1868governors, pontryagin1962ordinary}, defining small disturbances $\Delta \psi, \Delta \phi, \Delta \omega$ as
\begin{align*}
    \Delta \psi & \coloneqq \psi - \psi_0, \nonumber \\
    \Delta \phi & \coloneqq \phi - \phi_0, \nonumber \\
    \Delta \omega & \coloneqq \omega - \omega_0
    % \label{eq:differentials}
\end{align*}
% After specifying the load torque $G$ as \textcolor{red}{check second line in equation 4, is this really needed?}
% \begin{align}
%     G = k \cos (\psi_0)
%     \label{eq:loadTorque}
% \end{align}
After neglecting terms quadratic in disturbances $\Delta \psi, \Delta \phi, \Delta \omega$, we finally obtain
\begin{align}
    \Delta \dot{\psi} & = \Delta \phi \nonumber \\
    \Delta \dot{\phi} & = \frac{g \sin^2(\psi_0)}{l \cdot \cos(\psi_0)}\Delta \psi -\frac{b}{m} \Delta \phi + \frac{2 g \sin(\psi_0)} {l \cdot \omega_0} \Delta \omega \nonumber \\
    \Delta \dot{\omega} & = - \sin(\psi_0) \frac{k}{I} \Delta \psi
    \label{eq:WattGovernorFirstOrderLinearised}
\end{align}
These equations are typically used for the analysis of this engine-governor coupled system, and represent a simplified version of the governor's behaviour near equilibrium. The spindle angular velocity is initially assumed to be constant (by construction) to simplify the treatment from a spherical to a conical pendulum where the flyball arms' velocity is zero ($\phi = \phi_0 = 0$). Details about the steam engine are usually not provided, only explaining its effects through a torque $I \dot{\omega}$ which depends on the arm angle $\psi$ given a certain geometry of the governor, as here expressed in \eqnref{eq:Engine}.

\section{The governor's generative model}
Using active inference, we can re-derive similar equations, in particular for the engine's flywheel angular velocity $\Delta \dot{\omega}$ in \eqnref{eq:WattGovernorFirstOrderLinearised}, building on a previous formulation of PID control under this framework \citep{baltieri2019pid}. Our formulation includes a generative model in state-space form that describes observations/measurements, hidden states, inputs and parameters of the engine ``from the perspective of a governor''.

This description shouldn't however be taken too literally, as we will discuss later. In the spirit of \cite{van1995might, van1998dynamical}, we will rather highlight the somewhat bizarre nature of concepts such as \emph{measurements} performed by a physical system. In light of this, we thus stress an \emph{as-if} interpretation of generative models \citep{mcgregor2017bayesian, robert2007bayesian}. According to this idea, physical systems can be interpreted \emph{as if} they were cognitive agents, with generative models specifying their \emph{Bayesian beliefs}, governing preferences and dynamics that produce equations describing perception-action loops (see also Discussion). In our formulation, this corresponds to a rather unusual reading of the engine-governor coupled system: an agent trying to stabilise its observations, i.e., the perceived angle of the flyball arms, by adapting its own actions, i.e., the valve opening\footnote{Usually one looks at this system in terms of stabilising the velocity of the steam engine via the regulation of the arms angle, however the opposite perspective adopted here 1) currently better fits with the model presented in \cite{baltieri2019pid} and 2) is perfectly equivalent to the more traditional way of looking at this problem, since the angle $\psi$ is a single-valued monotonic function of the angular velocity $\omega$ \citep{pontryagin1962ordinary} due to the mechanics of the vertical shaft and the flyball arms, a ``stand-in'' in the sense of \cite{bechtel1998representations}. Interestingly, this view suggests that we might just as well consider the engine as an ``agent'' acting to control its observations of the ``environment'', the governor. One way to treat this issue in a more principled manner is to look at views of agency that are defined in terms of ``interactional asymmetry'' of the coupling between agent and environment \citep{barandiaran2009defining}. An in depth discussion of this idea is however left for future work.} (cf. ``behaviour as the control of perception'' \citep{powers1973behavior}). 

We thus start by defining the following generative model:
\begin{align}
    \psi & = x + z \nonumber \\
    x' & = - \alpha (x - \psi_0) + w
    \label{eq:StateSpaceWatt}
\end{align}
where $\psi$, $\alpha$, $x$, $x'$ and $\psi_0$ are, in state-space models terms, observations, parameters, hidden states and their derivatives\footnote{Here we denote derivatives in the generative model with a dash rather than a dot. The dynamics described by the generative model are not integrated directly, and are only used to derive a set of equations describing the \emph{recognition} dynamics as a gradient on variational free energy \citep{Friston2008c, buckley2017free} (see \eqnref{eq:GenericActionPerception}). The dot notation is used for the generative process in \eqnref{eq:WattGovernorFirstOrderLinearised}, and for the recognition dynamics \eqnref{eq:GenericActionPerception} (simplified, under a few assumptions, in \eqnref{eq:recognitionDynamicsFinal}), which would be forward-integrated in a simulation.}, and exogenous inputs of the engine from the (\emph{as-if}) perspective of the controller. Here the observations $\psi$ represent measurements performed by our agent, i.e., its incoming sensory input about the arm angle. Variables $x$, $x'$ define states (and their derivatives) of the engine hidden to the governor and assumed to generate these measurements (similar to the way one normally hides most details about the functioning of a real steam engine in \eqnref{eq:Engine}). Inputs $\psi_0$ stand for the presence of external factors affecting the system in the form of torque loads $G$, see \eqnref{eq:WattGovernorFirstOrderEquilibrium}. The first equation describes how hidden states $x$ are mapped to observations $\psi$ using an identity function, with a random variable $z$ introduced to express measurement noise, or rather uncertainty, from the perspective of our agent. In the second equation, the parameter $\alpha$ specifies the convergence rate of the intrinsic dynamics modelled by hidden states $x$ given inputs $\psi_0$. The fluctuations $w$ are introduced as an uncertainty term on the dynamics of the engine-governor coupled system, representing for instance errors due to the use of a linear approximation of the real dynamics near equilibrium.

Using this generative model, the \emph{recognition} dynamics of the system can be obtained by, 1) finding an expression for the variational free energy, and 2) under the assumption that the quantity of free energy is minimized over time, deriving a set of differential equations that minimize the free energy (i.e., following its negative gradient). Under a couple of assumptions described later, these differential equations (i.e., the recognition dynamics) reduce to the equation describing the dynamics of the linearised engine in \eqnref{eq:WattGovernorFirstOrderLinearised}.

An expression for the variational free energy can be derived after defining the distributions of $z, w$, in the simplest case assuming they are both Gaussian, $z \sim N (0, \sigma^2_z), w \sim N (0, \sigma^2_w)$
% \begin{align}
%     z \sim N (\mu_x, \sigma^2_z) \quad \quad \quad
%     w \sim N (\psi_0, \sigma^2_w)
%     \label{eq:RandomVariables}
% \end{align}
and by considering the following expression for the (Laplace-encoded) free energy \citep{Friston2008c}\footnote{For a full derivation of this particular form see for instance \cite{Friston2008a, buckley2017free, baltieri2019active}.}
\begin{align}
    F \approx - \ln P(\psi, x) \Bigr \rvert_{x = \mu_x}
\end{align}
where $F$ is evaluated near the most likely estimate of hidden states $x$, i.e., for a Gaussian distribution, its mean or median, $\mu_x$. After rewriting the generative model in \eqnref{eq:StateSpaceWatt} in probabilistic form using the definitions of $z, w$, one obtains the following expression
\begin{align}
    F \approx \frac{1}{2} \bigg[ \pi_{z} \Big( \psi - \mu_x \Big)^2 + \pi_{w} \Big (\mu'_x + \alpha(\mu_x - \psi_0) \Big)^2 - \ln{\big( \pi_{z} \pi_{w} \big)} \bigg]
    \label{eq:freeEnergy}
\end{align}
where we introduced precisions $\pi_z, \pi_w$ as the inverse variances of random variables $z, w$, i.e.,  $\pi_z = 1/\sigma_z^2, \pi_w=1/\sigma_w^2$. Actions $a$ are then defined under the very general assumption that they have an effect on observations $\psi$ \citep{Friston2010biocyb, baltieri2019active}, i.e.,
\begin{align}
    \psi = f(a)
    \label{eq:actionEffects}
\end{align}
By minimising free energy on both the means of hidden states $\mu_x$ and actions $a$, we then obtain the recognition dynamics \citep{buckley2017free}, i.e., a set of differential equations describing the dynamics of our agent. The recognition dynamics implement perception (estimation of states $\mu_x$) and action (control via actions $a$) of an agent described as a closed sensorimotor loop. The minimisation of free energy is then achieved via the following gradient descent
\begin{align}
    \dot{\tilde{\mu}}_x & = \tilde{\mu}'_x - k_p \frac{\partial F}{\partial \tilde{\mu}_x} \nonumber \\
    \dot{a} & = - k_a \frac{\partial F}{\partial a} = - k_a \frac{\partial F}{\partial \tilde{\psi}} \frac{\partial \tilde{\psi}}{\partial a}
    \label{eq:GenericActionPerception}
\end{align}
with learning rates $k_p$ and $k_a$. Notice that these equations use the same dot notation adopted to describe the generative process of the governor-engine coupled system in \eqnref{eq:WattGovernorFirstOrderLinearised} and represent an agent's dynamics actually integrated to implement (as-if) inference and control processes. Here we also introduced the use the tilde previously adopted by \cite{Friston2008c, buckley2017free, baltieri2019active} to represent higher embedding orders, in this case, $\tilde{\mu}_x = [\mu_x, \mu'_x ], \tilde{\psi} = [ \psi, \psi' ]$. More generally, when higher embedding orders are introduced as a possible way to represent non-Markovian processes \citep{Friston2008c, baltieri2019active}, this requires an extra term, $\tilde{\mu}'_x$, to ensure the convergence to a trajectory (rather than a point attractor) in a moving frame of reference \citep{Friston2008c}. For a Watt governor at equilibrium, we will however assume that the flyball arm angular velocity is zero, thus defining a point attractor where $\tilde{\mu}'_x = 0$. In this case, \eqnref{eq:GenericActionPerception} thus reduces to
\begin{align}
    \dot{\mu}_{x} & = \mu_x' - k_p \Big[ - \pi_z \big( \psi - \mu_{x} \big) + \pi_w \alpha \big(\mu'_{x} + \alpha(\mu_{x} - \psi_0) \big) \Big]  \nonumber \\
    \dot{\mu}'_{x} & = 0 \nonumber \\
    \dot{a} & = - k_a \frac{\partial \psi}{\partial a} \pi_z (\psi - \mu_{x})
    \label{eq:recognitionDynamicsInitial}
\end{align}
Under a few standard assumptions, this system can then be further simplified to show actions consistent with the regulation of the speed of a steam engine.

\subsection{Assumption 1: The dynamics of the generative model are overdamped}
The recognition dynamics in \eqnref{eq:recognitionDynamicsInitial} specify a gradient descent on the variational free energy in \eqnref{eq:freeEnergy} given the generative model in \eqnref{eq:StateSpaceWatt}. Importantly, this generative model is parameterized by $\alpha$, a parameter that describes the rate of convergence of its internal dynamics. As previously shown, for instance in \cite{baltieri2019active} (Chapter 7.), different choices of $\alpha$ allow for the implementation of qualitatively different behaviours: from the regulation of a process to a certain goal (large $\alpha$), to a purely passive (e.g., no actions) process of inference of the hidden properties of observed stimuli in the spirit of predictive coding models of perception \citep{rao1999predictive, baltieri2019dark} (small $\alpha$). Here we will assume a very large parameter $\alpha$ (i.e., $\alpha \gg 0$ and $\alpha \gg \pi_z, \pi_w$), leading to overdamped dynamics of the generative model in \eqnref{eq:StateSpaceWatt} dominated by the first (drift) term. This translates into recognition dynamics now describing updates of the average hidden state $\mu_x$ dominated by terms quadratic in $\alpha$ \citep{baltieri2019pid} (cf. \eqnref{eq:recognitionDynamicsInitial}). The expected hidden state $\mu_{x}$ thus quickly converges to its steady state (i.e., average velocity $\mu_x' = 0$), defined by the input (or bias term) $\psi_0$,
\begin{align}
    \dot{\mu}_{x} & \approx - k_p \pi_w \alpha^2 (\mu_{x} - \psi_0) \implies \mu_x = \psi_0
\end{align}
which readily ensures that the goal of the system is now to stabilise the flyball arms angle towards $\psi_0$ (NB: in general $\psi_0$ need not be a fixed point attractor), which in turn reflects regulation of the velocity of the engine\footnote{Under suitable parameters meeting typical stability criteria \citep{pontryagin1962ordinary}.} (seen also Note 1). Moreover, since we have assumed  $\alpha \gg \pi_z, \pi_w$, $\mu_x$ converges much more quickly than $a$ and thus the minimisation of free energy in \eqnref{eq:recognitionDynamicsInitial} can be further simplified to include only the equation for action
\begin{align}
    \dot{a} & \approx - k_a \frac{\partial \psi}{\partial a} \pi_z (\psi - \psi_0)
\end{align}

% \subsection{Assumption 1-bis: The dynamics are fully deterministic}
% It is important to differentiate the role of $\alpha$ and $\pi_w$ in case we want to give an interpretation in terms of Langevin dynamics for stochastic thermodynamics, i.e., the generative model as a colloidal particle. If we just assume large $\alpha$, according to Einstein relation $\pi_w$ should be small ($\sigma_w$ large). This is not good since that could invalidate assumption 1. To obviate that, it may suffice to assume a low temperature $T$ to satisfy the conditions to obtain $\pi_w \gg 0$ and $\alpha \gg 0$ at the same time.

\subsection{Assumption 2: Action updates are proportional to arm angle updates}
To show a direct correspondence between the active inference derivation and the original equations, here we assume that the measurement precision $\pi_z$ is inversely proportional to the moment of inertia of the steam engine flywheel $I$, 
\begin{align}
    \pi_z = \frac{1}{I} \implies \sigma_z^2 \propto I
\end{align}
or more precisely that $\pi_z = \frac{k}{I}$, using the constant $k$ relating flyball height and engine torque defined in \tabref{tab:variablesWattGovernor} \citep{pontryagin1962ordinary}. We then consider the case where the learning rate, a hyperparameter of the minimisation scheme (not of the generative model) is set to 1, $k_a = 1$, and essentially replaced by another hyperparameter playing a similar role, the precision $\pi_z$. We also assume a linear relationship between actions $a$ and observations $\psi$ in \eqnref{eq:actionEffects}, such that
\begin{align}
    \frac{\partial \psi}{\partial a} = \text{constant}
\end{align}
using the fact that $\partial \psi / \partial a$ need only be positive to ensure convergence via a negative feedback loop \citep{denny2002watt}. For convenience we impose $\partial \psi / \partial a = \sin(\psi_0)$, and finally obtain
\begin{align}
    \dot{a} & \approx - \sin(\psi_0) \frac{k}{I} (\psi - \psi_0) = - \sin(\psi_0) \frac{k}{I} \Delta \psi \color{red}{\Big( \equiv \Delta \dot{\omega} \Big)}
    \label{eq:recognitionDynamicsFinal}
\end{align}
which is equivalent to the simplified engine in \eqnref{eq:WattGovernorFirstOrderLinearised}, under the assumption that $\psi_0$ reflects changes due to different loads as in the original formulation of the governor-engine system \citep{pontryagin1962ordinary}. In active inference terms, \eqnref{eq:recognitionDynamicsFinal} describes the behaviour of an agent \emph{observing} its arms angle $\psi$ (thus indirectly \emph{inferring} the speed of the engine, see Note 2). At the same time, this agent \emph{acts} to produce a rotational energy of its flyballs based on deviations from the engine load torque $G$, and \emph{minimising} a prediction error between its observations $\psi$ and the load arm angle $\psi_0$ specifying an engine speed $\omega_0$ via a relation known as ``nonuniformity of performance'' \citep{pontryagin1962ordinary, denny2002watt}. Ultimately, this leads to a change in the engine's speed, here simplified as the angular velocity of the engine shaft (geared into the engine flywheel), $\Delta \omega$. Expected hidden states $\mu_x$ are only implicitly defined and effectively removed in the limit of deterministic dynamics using Assumption 1.

\section{Discussion}
% The centrifugal governor is a feedback system introduced in the late 18th century as a mechanism to control the speed of a steam engine, and popularised by James Watt. This regulator implements a negative feedback mechanism between a governor and a steam engine, the engine's velocity is controlled using a valve releasing steam, the governor regulates the opening of this valve via the accumulation of (angular) kinetic energy in a pair of rotating flyballs connected to a vertical shaft linked to the engine via a flywheel. When the kinetic energy is greater than the potential energy due to the gravitational acceleration, the valve is slowly closed, 

In this work, we introduced a rather unconventional treatment of a Watt governor based on active inference. The Watt (centrifugal) governor has played a central role in the debate between dynamicist and cognitivist ways of thinking about cognitive systems. Proposed as a dynamicist alternative analogy to the cognitivist digital computer \citep{van1995might} and the \emph{sense-model-plan-act} strategy implemented by several cognitivist frameworks \citep{brooks1991new, hurley2001perception}, since its introduction, a number of different works have argued for the merits and limitations of this analogy in addressing relevant questions in (philosophy of) cognitive science, including for instance \cite{eliasmith1997computation, bechtel1998representations, beer2000dynamical, chemero2009radical, seth2014cybernetic}. The discussion often focuses on the importance of representations to study and explain cognitive agents, echoing a long standing debate over the role of these constructs for theories of cognition \citep{fodor1983modularity, harvey1992untimed, varela1991embodied, beer2000dynamical, gallagher2006body, chemero2009radical, di2017sensorimotor}. Echoing \cite{harvey1992untimed} and following in particular \cite{chemero2009radical}, we distinguish between metaphysical and epistemological claims when it comes to representationalism. The first pertain to the nature of cognitive systems, the second to our (the scientists') best explanation of cognitive systems. In the case of the Watt governor, one is hard pressed to defend the metaphysical claim. Rather, the debate usually focuses on the epistemological status of the system: is it useful to explain a Watt governor in representational terms or, in other words, does taking the governor flyball arm angle to \emph{represent} the speed of the engine help us understand the workings of the governor?

Active inference is a recent framework developed in theoretical neuroscience that describes several aspects of living and cognitive systems in terms of the minimisation of variational (or expected) free energy, which under simplifying (Gaussian) assumptions reduces to a weighted sum of prediction errors. According to active inference, action-perception loops can be seen as a gradient descent on a free energy functional describing normative behaviour for a system. Perception is characterised as a process of Bayesian inference on hidden world variables and action is portrayed as the update of environmental properties to better reflect current perceptual inferences, mediated by the goals of an organism, e.g., its survival, in the form of prior Bayesian beliefs. It has been argued that this framework, often addressed in terms of predictive processing, constitutes a new paradigm for the study of cognitive agents \citep{clark2013whatever, hohwy2013predictive, seth2014cybernetic, clark2015surfing, wiese2017vanilla, hohwy2020new}. At the moment however, its role in cognitive science remains highly controversial \citep{kirchhoff2017there, colombo2018realist}. In particular, some authors suggest that internal representations are central to predictive processing and are used to define computational processes in a more classical sense \citep{hohwy2013predictive, hohwy2020new}, some others claim that representations are detrimental for a proper understanding of predictive processing \citep{bruineberg2018anticipating}, while others attempt to reconcile these different interpretations \citep{seth2014cybernetic, clark2015surfing}. Much of this literature, however, seems to remain (almost deliberately) unclear as to the metaphysical status of the central constructs involved, with a few relevant works explicitly favouring an instrumentalist interpretation \citep{anderson2017bayes, colombo2017explanatory, colombo2018realist, van2020living}.

Here we take a rather sobering perspective on the role of active inference theories for cognitive and living systems, especially arguing against claims regarding internal representations in predictive processing and the metaphysical status often implicitly granted to processes of variational free energy/prediction error minimisation \citep{friston2013life, clark2013whatever, hohwy2013predictive, gladziejewski2018predictive, friston2019free}. Our perspective aligns with epistemological stances \citep{robert2007bayesian, beer2015information} supporting an instrumental role for frameworks such as Bayesian decision theory, information theory, dynamical systems, and in this case active inference and predictive processing \citep{colombo2017explanatory}, as possible \emph{interpretations} of physical and cognitive systems. In this light, a generative model may simply provide a possible way to describe a system's behaviour, rather than an ontological characterisation of its very nature and inner workings \citep{baltieri2019generative}. Generative models can act as representations for an observer, as a valuable tool for the study of physical, living and cognitive systems, to describe them from an experimenter's perspective \citep{bechtel1998representations, harvey2008misrepresentations, chemero2009radical, beer2015information, mcgregor2017let}, e.g., to understand the relationship between the spindle arm angle in a Watt governor, and the speed of a steam engine. Their interpretation as intrinsic properties of a system, i.e., generative models as ``internal models'' used by the system itself \citep{fodor1983modularity}, is on the other hand misleading.

Should one try to find a generative model of the kind expressed in \eqnref{eq:StateSpaceWatt} -- inside -- a Watt governor? No, as this is simply a category mistake: observer-dependent uncertainties (e.g., $w, z$) and arbitrary assumptions (e.g., parameter $\alpha \gg 0$) are relational terms that cannot be found \emph{within} this physical system \citep{harvey2008misrepresentations}. A generative model provides details for a scientist to specify a cost functional, variational free energy, that can be used to describe the (recognition) dynamics of a system \citep{buckley2017free, ramstead2019tale} as seen in \eqnref{eq:recognitionDynamicsInitial}. The presence of a ``generative model for a Watt governor'' in \eqnref{eq:StateSpaceWatt}, and the ensuing claims of an agent ``minimising variational free energy'' while regulating the opening of a steam valve, should thus be handled with care\footnote{In the same way one should be careful when explaining, for example, Coulomb's law: saying that an electron \emph{calculates} its distance from another electron, \emph{computes} the force applied to that electron and \emph{actuates} said force in the real world would make for a rather unusual and possibly confusing explanation \citep{latash2010two}.}.

This suggests that we should exercise caution in making metaphysical claims using predictive processing, the free energy principle and active inference. Statements regarding their mechanistic \citep{clark2013whatever, hohwy2013predictive, hohwy2020new} or representational \citep{rescorla2016bayesian, hohwy2016self} contents for cognitive systems are in fact often ambiguous and mostly based only on ``evidence consistent with'' generative models in the brain \citep{colombo2017explanatory, colombo2018realist}, as even the most recent reviews show \citep{walsh2020evaluating}.

At this point one might wonder what the relevance of active inference and predictive processing might be in the cognitive sciences. Here we suggest that probabilistic generative models ought to be recognised for their effectiveness as a mathematical formalism connecting and extending ideas such as the good regulator theorem in cybernetics \citep{conant1970every} (see \cite{seth2014cybernetic}), the internal model principle in control theory \citep{francis1976internal} (see \cite{baltieri2019active}), perceptual control theory \citep{powers1973behavior} in psychology, or the notion of entailment in theoretical biology \citep{rosen1991life} (see \cite{friston2012free, ramstead2019tale}). In this light, active inference can play an important instrumental role for the overarching attempt of unifying previous results in the study of adaptive agents in cybernetics, control theory, psychology, neuroscience, dynamical systems, information theory and physics. Under a more general framework, notions such as feedback, stability, inference, attractors, uncertainty and dynamics can be seen from different, but complementary, perspectives that allow for more complete descriptions of agents and agency, similarly to the approach adopted, for instance, by \cite{beer2015information}. 

For example, while the notion of ``inference'' in cognitive science often hinges on the intuition of (passive) \emph{inference to the best explanation} \citep{chemero2009radical} and its limited explanatory power for behaviour and agency \citep{seth2015inference, baltieri2019active}, a more general treatment of Bayesian inference schemes and their connections to different fields can extend this narrow notion to include more useful interpretations of inference. These include connections to stochastic processes in non-equilibrium thermodynamics \citep{seifert2012stochastic}, where biological agents can be better characterised as systems away from (thermodynamic) equilibrium \citep{schrodinger1944life}. In this case, inference becomes simply a way to (mathematically) describe the statistical properties of agents represented in terms of non-equilibrium steady states of a random process \citep{friston2019free}. A control theoretic reading of inference \citep{kappen2012optimal} then introduces the idea of \emph{inference to the most useful (normative) behaviour} \citep{baltieri2019active, seth2015inference}, rather than \emph{to the best explanation}. Here, actions and policies of an agent are described as an inferential process \emph{biased} towards ensuring that its normative constraints are met (i.e., a Bayesian inference process \emph{conditioned} on a system's norms, such as its survival and very existence \citep{barto2013novelty, baltieri2019active}). Furthermore, one can then also recognise inference as a relevant description of theories in evolutionary biology, where following adaptive paths on the fitness landscape is for example likened to a problem of Bayesian model selection \citep{czegel2019multilevel}.

\section{Conclusion}
Using the Watt governor as a toy model, in this work we discussed the importance of generative models in the context of predictive processing and active inference, extending our previous critiques on their role as internal representations \citep{bruineberg2018anticipating, baltieri2017active, baltieri2019generative}. By defining a linear probabilistic generative model describing ``observations'' and ``hidden states'' from the perspective of a Watt governor, we built a cost functional (i.e., variational free energy) to describe the behaviour of this system \emph{as if} \citep{mcgregor2017bayesian} it was an agent trying to minimise its prediction error to control its observations and regulate the speed of a steam engine. Via the minimisation of this cost functional under a couple of relatively straightforward and common assumptions, i.e., overdamped dynamics and an appropriate use of constants that ensure regulation via a negative feedback loop, we then re-derived equations equivalent to the mathematical treatment of an engine for the Watt governor linearised near equilibrium. Using this as an example, we then discussed epistemological and metaphysical stances \citep{chemero2009radical} for generative models in predictive processing and active inference, focusing on the former and exerting caution on the latter. Our formulation shows that generative models can easily be used to describe canonical cases of the dynamical systems approach in cognitive science. 

This paper suggests that generative models are best understood as an epistemic tools for an observer to specify the properties of a system they wish to study and their own assumptions and sources of uncertainty during a process of epistemological analysis \citep{colombo2018realist, van2020living}. As such, they are not \emph{internal} representations for a system, but rather constitute just a descriptive framework (a representation, not internal \citep{harvey2008misrepresentations}) for an observer. Seeing generative models as \emph{internal} representations may simply reflect a ``mind projection fallacy'' \citep{jaynes1990probability}, where epistemic constructs used by a scientist are assumed to be real objects in the physical world.

Within the existing literature on the free energy principle, active inference and predictive processing, we find an almost deliberate conflation of realist and instrumentalist perspectives, addressing how aspects of one's model come to explain or constitute aspects of the mind. One example is the discussion on boundaries of the mind (presumably aspects of the world), which is premised on \emph{where} Markov Blankets (i.e., statistical properties of a model specifying conditional independence between random variables) are located \citep{clark2017knit, hohwy2017entrain, kirchhoff2019determine}. This case will however be addressed in more detail in the near future. 

The present work supports active inference as a potentially useful descriptive language for cognition, highlighting its instrumental role in studying action-perception within a general mathematical framework including complementary interpretations of an agent's behaviour \citep{tishby2011information, beer2015information}, but remains cautious on the metaphysical implications of generative models as internal representations often found in the literature \citep{hohwy2013predictive, clark2015surfing, rescorla2016bayesian}.

\section{Acknowledgments}
MB is a JSPS International Research Fellow supported by a JSPS Grant-in-Aid for Scientific Research (No. 19F19809). CLB was supported in part by a BBSRC Grant BB/P022197/1. JB was supported by a Macquarie University Research Fellowship. MB wishes to thank Inman Harvey and Filippo Torresan for insightful discussions that helped improving different aspects of this work.

\footnotesize
\bibliographystyle{apalike}
\bibliography{AllEntries} % replace by the name of your .bib file

\end{document}